\documentclass[aps,pra,twocolumn,groupedaddress,showpacs,floatfix]{revtex4}

\usepackage{graphicx}

\begin{document}


\title{Simplified motional heating rate measurements of trapped ions}


\author{R. J. Epstein}
\email[]{rje@nist.gov}
\author{S. Seidelin}
\author{D. Leibfried}
\author{J. H. Wesenberg}
\author{J. J. Bollinger}
\author{J. M. Amini}
\author{R. B. Blakestad}
\author{J. Britton}
\author{J. P. Home}
\author{W. M. Itano}
\author{J. D. Jost}
\author{E. Knill}
\author{C. Langer}
\altaffiliation{Present address: Lockheed Martin, Huntsville, Alabama, USA}
\author{R. Ozeri}
\altaffiliation{Present address: Weizmann Institute of Science, Rehovot, Israel}
\author{N. Shiga}
\author{D. J. Wineland}
\affiliation{National Institute of Standards and Technology, Boulder, Colorado 80305, USA}



\date{\today}

\begin{abstract}
We have measured motional heating rates of trapped atomic ions, a factor that can influence multi-ion quantum logic gate fidelities. Two simplified techniques were developed for this purpose: one relies on Raman sideband detection implemented with a single laser source, while the second is even simpler and is based on time-resolved fluorescence detection during Doppler recooling. We applied these methods to determine heating rates in a microfrabricated surface-electrode trap made of gold on fused quartz, which traps ions 40~$\mu$m above its surface. Heating rates obtained from the two techniques were found to be in reasonable agreement. In addition, the trap gives rise to a heating rate of $300 \pm 30$~s$^{-1}$ for a motional frequency of 5.25~MHz, substantially below the trend observed in other traps.
\end{abstract}


\maketitle


\section{Introduction}

Control of the quantum states of trapped ions has progressed significantly over the past few years. Many of the necessary requirements for quantum information processing have been demonstrated in separate experiments, such as high-fidelity state preparation, read-out, single- and two-qubit gates, and long-lived single-qubit coherence (See e.g. Ref.~\cite{daveICAP}). One of the limitations thus far in scaling to larger numbers of ions has been the lack of a suitable trap architecture. A critical benchmark for a trap design is the heating rate of an ion's motional degrees of freedom due to electric field noise from the trap electrodes. As current quantum gates rely on the coupled motion of two or more ions, noise in the motion can degrade gate fidelities~\cite{sorensen}. To facilitate the determination of these heating rates, we have developed two measurement methods that have reduced hardware complexity compared to that of more traditional methods~\cite{diedrich,monroe,meekhof,roos,tamm,turch,rowe,devoe,desla04,roee,stick,desla,jonathan,ike}.  Here, we report details of these two methods.

The ion trap used for this study is a monolithic design made of gold on fused quartz, where all trap electrodes reside in a single plane~\cite{chiav, signe}. This ``surface-electrode" geometry has the potential to greatly simplify the trap fabrication process and electrode wiring topology, thereby enabling the creation of large multiplexed trap arrays. Heating rates were previously measured in a nearly identical trap by recording time-resolved fluorescence during Doppler recooling after allowing the ions to heat up~\cite{signe}. The details of this technique are examined in a recent theoretical paper~\cite{janus}. This method was relatively simple to implement and the measured rates were promisingly low. However, the accuracy of the technique was uncertain because it relies on changes in vibrational quanta of order 10$^4$, whereas quantum gate fidelities can depend on changes of a single quantum. It was an open question whether the heating rates could be reliably extrapolated down to the single quantum level. 

To test the accuracy of the Doppler recooling method, we have built a simplified Raman sideband cooling apparatus and measured heating rates of one degree of freedom in the single quantum regime. We also measured heating rates using Doppler recooling under similar experimental conditions, and find reasonable agreement between the values obtained with the two techniques. In addition, heating rates were measured at several trap frequencies and the electric field noise was found to have approximately 1/f character in this particular trap. Finally, although the trap discussed here was fabricated by the same process as that of Ref.~\cite{signe}, the heating rates are found to be somewhat lower, possibly due to cleaner electrode surfaces~\cite{trapnote}.

\section{\label{sec:recool}Doppler recooling}

The Doppler recooling method is based on the observation that the near-resonance fluorescence rate from an ion is influenced by its motional temperature due to the Doppler effect. By monitoring the fluorescence as a function of time during Doppler cooling of an initially hot ion, one can determine the initial temperature of the ion averaged over many experimental runs. In the experiments discussed here, an ion is first cooled close to the Doppler limit. Then it is allowed to heat up for a variable amount of time (the delay time) by turning off the Doppler cooling laser beam. The laser is turned back on and the fluorescence is monitored as a function of time until the ion's fluorescence rate reaches its steady-state value. By fitting a theoretical model~\cite{janus} to the data, the ion's temperature at the end of the delay time can be extracted. The model is a one-dimensional semi-classical description of Doppler cooling in the ``weak-binding" limit, where the ion's motional frequency is much smaller than the linewidth of the Doppler cooling transition (See Ref.~\cite{janus} for details). An attractive feature of this technique is its relative simplicity. It requires only one low-power red-detuned laser beam and no magnetic fields~\cite{signe}, in contrast to the Raman sideband technique discussed in Section~\ref{sec:raman}.

\begin{figure}
\includegraphics{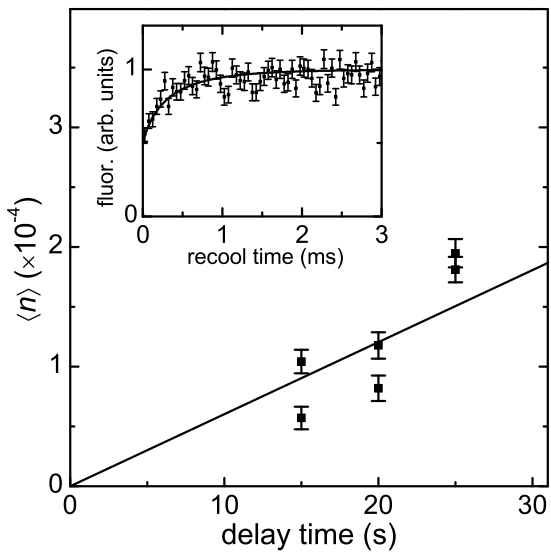}
\caption{\label{fig:recool} Average number of vibrational quanta $\langle n\rangle$ as a function of delay time obtained by the Doppler recooling method with an axial trap frequency of 4.02~MHz. The fit (solid line) gives a heating rate of $d\langle n\rangle/dt = 620 \pm 50$~s$^{-1}$. Inset: Ion fluorescence versus recooling time for a delay time of 25~seconds with fit (solid line) to the model of Ref.~\cite{janus}.}
\end{figure}

Figure~\ref{fig:recool} displays the average number of axial vibrational quanta $\langle n\rangle$ for various delay times obtained from Doppler recooling measurements. The axial trap frequency was $\omega=2\pi\times4.02$~MHz. Here the axial direction is the direction of weakest binding in the trap and is controlled primarily by static potentials~\cite{signe}. Each value of $\langle n\rangle$ was obtained by fitting the model of Ref.~\cite{janus} to a Doppler recooling trace, as exemplified by the inset data. The model yields the thermal energy of the ion at the start of recooling ($\approx \hbar \omega\langle n\rangle$), which was then converted to vibrational quanta. A weighted linear fit to $\langle n\rangle$ versus delay time yields a heating rate of $d\langle n\rangle /dt = 620 \pm 50$~s$^{-1}$. The fit was constrained to pass through the origin because the ion was initially cooled near the Doppler limit of a few quanta. We used the reciprocal of the estimated variances as the weights in all of the weighted fits presented here. We note that the quoted uncertainties include estimated statistical uncertainties only~\cite{varnote}.

To obtain these data, magnesium ions were created in the trap through a two-photon photoionization process using 1 to 10~mW of 285~nm laser excitation~\cite{madsen}. This ionization method was found to significantly reduce the required Mg oven temperature (and the concomitant pressure rise) compared to electron impact ionization. All experiments were carried out with $^{25}$Mg$^+$ ions in a magnetic field of 10~G for consistency with the Raman measurements discussed below. In addition, the Doppler beam (``Blue Doppler") saturation parameter was 0.9 throughout. Due to the hyperfine structure of $^{25}$Mg$^+$, a second laser beam (``Red Doppler") was used to repump out of the $F = 2$ ground state manifold. This additional beam was not necessary in the measurements of Ref.~\cite{signe} performed at $B\approx 0$ with $^{24}$Mg$^+$, an isotope without hyperfine structure. These beams have the same intensities, polarizations and detunings as in the Raman experiments described below (See Figs.~\ref{fig:beams} and~\ref{fig:ramansb}).

\section{\label{sec:raman}Simplified Raman Sideband Detection}

\begin{figure}[b]
\includegraphics{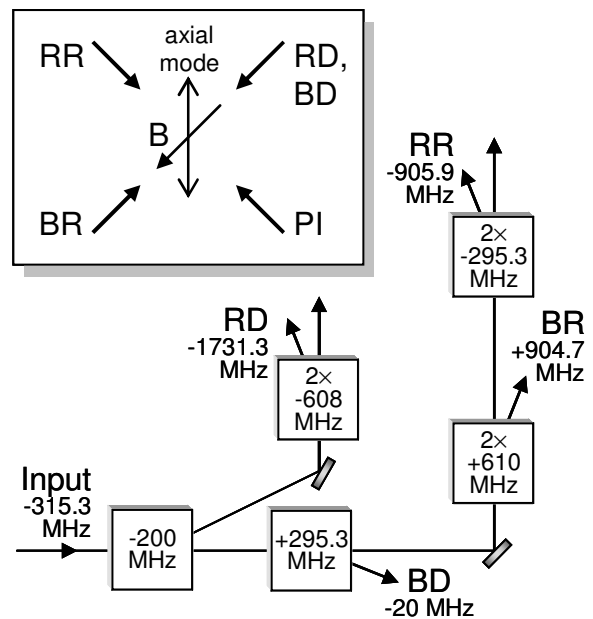}
\caption{\label{fig:beams} Schematic of the laser beams. All beams are derived from one source by use of acousto-optic modulators (AOMs) as frequency shifters; they are labeled Red Raman (RR), Blue Raman (BR), Red Doppler (RD), and Blue Doppler (BD). The AOMs (boxes) are labeled by the frequency shift they impart to the deflected beams; 2$\times$ indicates double-pass configuration. The frequency shift is noted for each beam relative to the $^2S_{1/2},|3,-3\rangle$ $\leftrightarrow$ $ ^2P_{3/2},|4,-4\rangle$ cycling transition. Inset: Geometry of beams, trap axis and magnetic field $B$, including the photoionization (PI) beam.}
\end{figure}

\begin{figure}
\includegraphics{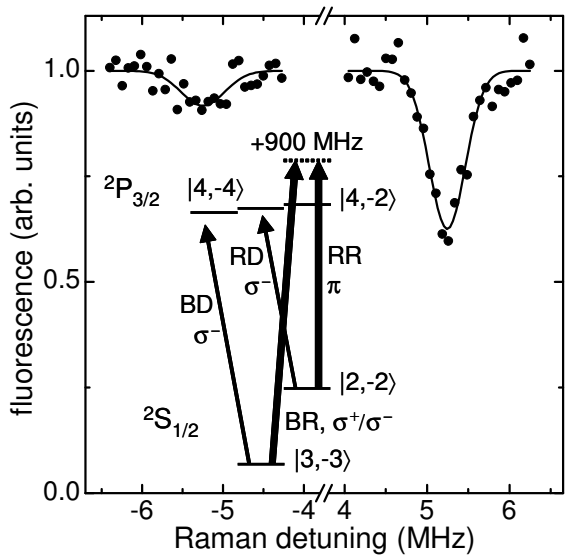}
\caption{\label{fig:ramansb} Ion fluorescence as a function of Raman frequency detuning relative to the Raman carrier ($\Delta n = 0$) transition. Fits (solid lines) to the Raman sideband amplitudes yield $\langle n\rangle = 0.34 \pm 0.08$, the average number of axial vibrational quanta after Raman cooling with a trap frequency of 5.25~MHz. Inset: Partial level diagram of $^{25}$Mg$^+$ showing the laser beams and their polarizations. Relevant $|F,m_F\rangle$ levels are indicated. For $^{25}$Mg$^+$ in a magnetic field of 10~G, the $^2S_{1/2}$, $|3,-3\rangle$ and $|2,-2\rangle$ levels are split by $1.81$~GHz. The $^{2}S_{1/2}$ $\leftrightarrow$ $^{2}P_{3/2}$ transition wavelength is 280~nm.}
\end{figure}

Our Raman sideband detection apparatus has been simplified compared to more commonly used schemes~\cite{diedrich,monroe,meekhof,roos,tamm,turch,rowe,devoe,desla04,roee,stick,desla,jonathan,ike}. Instead of relying on three or more laser sources, Fig.~\ref{fig:beams} depicts how the two Raman beams and two Doppler cooling beams were derived from a single 280~nm source, a frequency-quadrupled fiber laser. The frequency-doubled output of the laser was frequency-locked to an iodine vapor absorption line. Multiple acousto-optic modulators (AOMs) were used as frequency shifters and on/off switches, controlling the beams that we call Red Raman (RR), Blue Raman (BR), Red Doppler (RD), and Blue Doppler (BD). Referring to the level diagram of $^{25}$Mg$^+$ in the inset of Fig.~\ref{fig:ramansb}, a scheme relying on multiple lasers would typically employ one laser to drive the $^2S_{1/2}$ $\leftrightarrow$ $^2P_{1/2}$ transition (not shown) for optical pumping, a second laser to drive the $^2S_{1/2}$ $\leftrightarrow$ $^2P_{3/2}$ transition for state detection (on a cycling transition), and a third laser for far-off-resonant Raman beams~\cite{monroe}. By contrast, only a single laser is needed if the $^2S_{1/2}$ $\leftrightarrow$ $^2P_{3/2}$ transition is used for Doppler cooling, state preparation and detection (with possibly reduced state preparation fidelity) and if we accept relatively low Raman detuning. The double-passed AOMs in Fig.~\ref{fig:beams} generate two beams for Raman transitions with an adjustable frequency difference near 1810~MHz and detunings from the $^{2}P_{3/2}$, $|4,-2\rangle$ state of approximately 900~MHz as shown in the inset of Fig.~\ref{fig:ramansb}. This relatively small Raman detuning, given the optical transition linewidth of 41.4~MHz, leads to significantly reduced coherence of Raman transitions through incoherent photon scattering~\cite{roee}. In particular, the Rabi flopping decay time for the red sideband ($n\rightarrow n-1$) Raman transition~\cite{meekhof} is approximately one Rabi oscillation period after sideband cooling to $\langle n\rangle\approx 1$. Despite these compromises, we are able to achieve reasonable sideband detection contrast~\cite{polnote} and cool the axial mode to $\langle n\rangle = 0.34 \pm 0.08$~quanta for an axial trap frequency of 5.25~MHz (Fig.~\ref{fig:ramansb}), which is sufficient for heating measurements.

\begin{figure}
\includegraphics{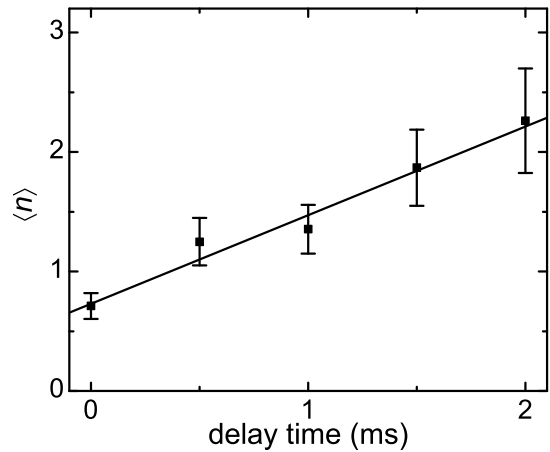}
\caption{\label{fig:raman} Average number of vibrational quanta $\langle n\rangle$ as a function of Raman measurement delay time for an axial trap frequency of 4.02~MHz. The fit (solid line) gives a heating rate of $d\langle n\rangle/dt = 690 \pm 60$~s$^{-1}$ (averaged over multiple similar data sets).}
\end{figure}

\begin{figure}
\includegraphics{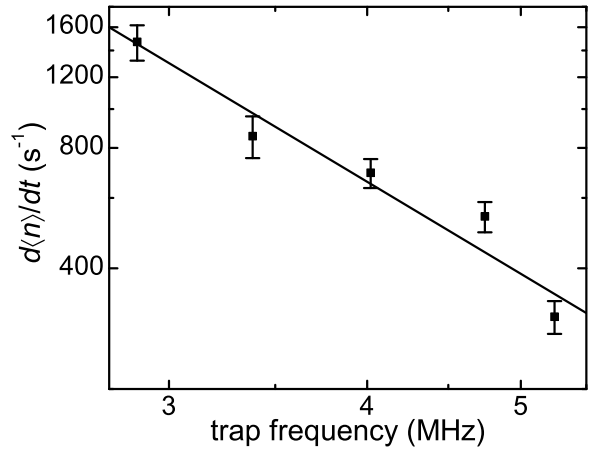}
\caption{\label{fig:hrvsf} Heating rate $d\langle n\rangle/dt$ as a function of axial trap frequency. The lowest measured heating rate is $300 \pm 30$~s$^{-1}$ at 5.25 MHz. The fit (solid line) gives a frequency dependence of  $d\langle n\rangle/dt \propto \omega^{-2.4 \pm 0.4}$.}
\end{figure}

\begin{figure*}
\includegraphics{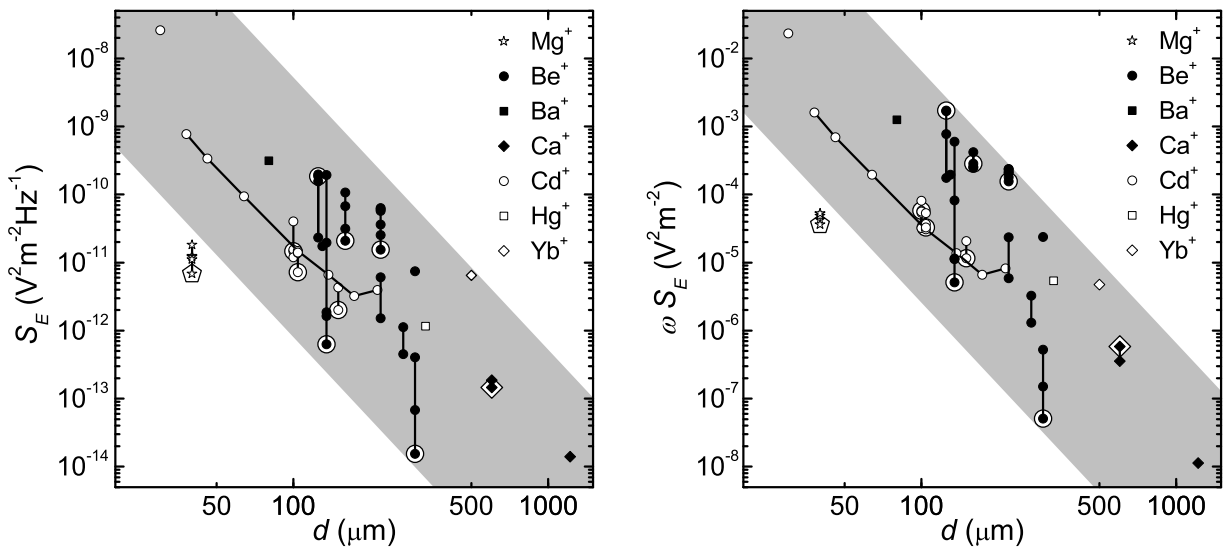}
\caption{\label{fig:noise} Electric field noise spectral density $S_E$ (left panel) and $\omega S_E$ (right panel) for traps with varying distance $d$ between the ion and the closest electrode. Data for the same trap are connected by line segments; the data point corresponding to the largest value of $\omega$ (if not constant) is marked by a larger symbol with a border. The gray bands depict $d^{-4}$ scaling. References for data are: Mg$^+$ (this work), Ba$^+$ \cite{devoe}, Be$^+$ \cite{monroe, turch, rowe}, Ca$^+$ \cite{roos, jonathan}, Cd$^+$ \cite{desla04, desla, stick}, Hg$^+$ \cite{diedrich}, and Yb$^+$ \cite{tamm}.}
\end{figure*}

In a typical experimental sequence, we first perform Doppler cooling and state preparation with BD and RD for 300~$\mu$s followed by BD for 20~$\mu$s and then RD for 20~$\mu$s. Then we apply 25 to 30 cycles of resolved sideband cooling~\cite{monroe}. One cycle consists of an approximate $\pi$-pulse on the red Raman sideband transition ($n\rightarrow n-1$, $n\approx 1$) using RR and BR, followed by RD for 8~$\mu$s and then BD for 0.3~$\mu$s. These pulses typically enable cooling of the ion to $\langle n\rangle \leq 1$ axial vibrational quanta. In order to measure heating rates, all beams are turned off for a specified delay period (usually 0 to 5~ms) to let the ion heat up. Next, a Raman pulse is applied with variable BR frequency-detuning; the pulse time is chosen such that a $\pi$-pulse is effected when resonant with the red sideband transition ($n\rightarrow n-1$, $n\approx 1$). Finally, ion fluorescence is detected with BD for 50~$\mu$s. The sequence is repeated for different BR detunings to sweep out the sidebands, as in Fig.~\ref{fig:ramansb}, where each data point is typically an average of several hundred experiments (1400 for the data in Fig.~\ref{fig:ramansb}). The Rabi flopping $\pi$ time on the red sideband was 1 to 5~$\mu s$, depending on the axial trap frequency and laser beam intensities. Typical laser powers were 1~$\mu$W (BD), 10~$\mu$W (RD), 90~$\mu$W (BR), and 40~$\mu$W (RR); beam waists were estimated to be 15 to 30~$\mu$m.


\section{Technique comparison and surface trap results}

Figure~\ref{fig:raman} shows values of $\langle n\rangle$ extracted from Raman sideband measurements at multiple delay times for the same 4.02~MHz axial trap frequency as Fig.~\ref{fig:recool}. Each value of $\langle n\rangle$ is obtained from Gaussian fits to the sidebands as in Fig.~\ref{fig:ramansb}. Assuming $n$ has a thermal distribution, then $\langle n\rangle = \frac{R}{1-R}$, where $R$ is the ratio of the red and blue sideband amplitudes~\cite{monroe, turch}. A weighted linear fit to $\langle n\rangle$ versus delay time yields a heating rate of $d\langle n\rangle /dt = 690 \pm 60$~s$^{-1}$. This compares reasonably well with the value obtained from Doppler recooling (See Section~\ref{sec:recool}). Likewise, at a different trap frequency of 2.86~MHz, we find heating rates of $1470 \pm 150$~s$^{-1}$ and $1260 \pm 130$~s$^{-1}$ for Raman and Doppler recooling techniques, respectively.

Figure~\ref{fig:hrvsf} displays heating rates, for a range of axial trap frequencies, measured with the Raman sideband technique. A weighted power-law fit yields $d\langle n\rangle/dt \propto \omega^{-2.4 \pm 0.4}$. From the heating rates we can calculate the electric field noise spectral density $S_E(\omega)\approx \frac{d\langle n\rangle}{dt}\frac{4m\hbar\omega}{e^2}$, with ion mass $m$, and electron charge $e$~\cite{turch}. Given the explicit factor of $\omega$ in this equation and the measured frequency dependence of $d\langle n\rangle/dt$, we find $S_E \propto \omega^{-1.4 \pm 0.4}$ for our surface-electrode trap. A similar frequency dependence has been observed in ion traps of different geometries~\cite{turch,desla04,desla} and may give some insight into the heating mechanism.

In Fig.~\ref{fig:noise}, we put these heating results in perspective by plotting values of $S_E(\omega)$ and $\omega S_E(\omega)$ versus $d$, the distance between the ion and the nearest electrode, for several different ion traps reported in the literature (as similarly done in Ref.~\cite{desla04}). The surface trap studied here has $d=40$~$\mu$m. For comparison, all the traps plotted have approximately room temperature electrodes; it has recently been found that cooling the electrodes can significantly reduce the heating rates~\cite{desla,ike}. While the fundamental heating mechanism is not understood, the predominant explanation is that the electrodes are covered in patches of varying potential that fluctuate with an unknown frequency dependence. If we assume that these fluctuating patch potentials have a size $\ll d$, then $S_E$ should scale as $d^{-4}$ (indicated by the gray shaded bands) \cite{turch}. A similar dependence on $d$ was observed in Ref.~\cite{desla}, where $d$ could be varied in the same trap (Fig.~\ref{fig:noise}, open circles).

Concerning the frequency spectrum of the noise, if $S_E \propto 1/\omega$, then $\omega S_E$ would of course be independent of $\omega$. Fig.~\ref{fig:noise} shows several cases where the values of $\omega S_E$ for a given trap are bunched together, indicating that $S_E\propto 1/\omega$ is a better assumption than $S_E$ being independent of $\omega$. In most traps shown, however, $S_E$ actually depends more strongly on $\omega$. It is unclear whether this is intrinsic to the traps or due to external noise sources.

As can be seen, the values of $S_E$ and $\omega S_E$ for the NIST surface trap are over an order of magnitude lower than what might be expected from the trend. The significant scatter in the data points highlights the importance of other parameters that have yet to be fully quantified, such as the microscopic properties of the electrodes (purity, roughness, crystallinity, etc.), which in turn may depend on processing procedures. For example, there is some evidence that electrode contamination (due to the ion loading process) influences the electric field noise~\cite{turch}. In our apparatus, the loading geometry is such that the electrode surfaces become coated with a small amount of Mg during each loading attempt. While we have not measured a systematic change in the heating rate due to loading in this trap, we cannot rule out the influence of surface contamination.

\section{Conclusion}

According to the results presented here, the simple Doppler recooling technique is a reasonably accurate tool for trap characterization. It has several advantages, including simplicity and relatively small resource requirements (a single low-power laser, no magnetic fields, etc.), which results in lower cost and setup time. The primary disadvantage is that delay times can be inconveniently long for low heating rates. For example, delay times of approximately one minute and averaging durations of several hours were needed for a heating rate of 300~s$^{-1}$ with our experimental parameters. This would be particularly troublesome if the uncooled ion lifetime in the trap (set by background gas collisions) were comparable to the delay time. However, weighing these factors, the Doppler recooling technique may still prove useful for rapid characterization of new ion traps. Means to potentially reduce the measurement time for this method are outlined in Ref.~\cite{janus}.

Accepting somewhat increased complexity, a simplified Raman sideband detection apparatus is shown to be suitable for heating rate measurements in the single quantum regime. Using relatively small Raman beam detunings (900~MHz) and the same ($^2S_{1/2}$ $\leftrightarrow$ $^2P_{3/2}$) transition for Doppler cooling, state preparation, and detection, enables a single laser to supply all necessary beams. Despite reduced sideband cooling efficiency and Rabi coherence, cooling of a single mode to $\langle n \rangle \leq 1$ is achieved with significantly fewer resources than more common Raman sideband detection experiments~\cite{diedrich,monroe,meekhof,roos,tamm,turch,rowe,devoe,desla04,roee,stick,desla,jonathan,ike}.

Finally, heating rates for the new surface-electrode trap geometry~\cite{signe} appear to be manageable for large-scale quantum information processing. Compared to other ion traps (Fig.~\ref{fig:noise}), the rates measured here are significantly lower than one might expect for the electrode--ion separation in this trap. A future surface-electrode trap design---having multiple zones with differing values of $d$ and a loading scheme that does not contaminate the electrodes---would be useful to further characterize the scaling of the electric field noise.

\begin{acknowledgments}
Work supported the Disruptive Technology Office (DTO) under contract number 712868 and NIST. R.J.E. and J.M.A. acknowledge National Research Council Research Associateship Awards. S.S. acknowledges support from the Carlsberg Foundation. J.H.W. acknowledges support from The Danish Research Agency. J.P.H. acknowledges support from a Lindemann Fellowship. 

This manuscript is a publication of NIST and not subject to U.S. copyright.
\end{acknowledgments}


\begin{thebibliography}{23}
\bibitem[{tra()}]{trapnote}
\bibinfo{note}{The trap of Ref.~\cite{signe} had significant magnesium
  contamination on the electrodes.}
\bibitem[{var()}]{varnote}
\bibinfo{note}{The measured heating rates varied from run to run by somewhat more than expected from the estimated uncertainties of a given run for unknown reasons. Quoted heating rates are averages of data taken on different runs.}
\bibitem[{pol()}]{polnote}
\bibinfo{note}{The optical windows were found to have spatially inhomogeneous
  birefringence leading to reductions in both optical polarization purity and
  atomic state preparation fidelity.}
\end{thebibliography}

\end{document}